# Interplay of canted antiferromagnetism and nematic order in Mott insulating $Sr_2Ir_{1-x}Rh_xO_4$

Hyeokjun Heo[1,2,*], Jeongha An[1,2,*], Junyoung Kwon[3], Kwangrae Kim[3], Youngoh Son[1], B. J. Kim[3], Joonho Jang[1,2,†]

[1]*Department of Physics and Astronomy and Institute of Applied Physics, Seoul National University, Seoul 08826, South Korea*

[2]*Center for Correlated Electron Systems, Institute for Basic Science, Seoul 08826, South Korea*

[3]*Department of Physics, Pohang University of Science and Technology, Pohang 37673, South Korea*

**ABSTRACT**

$Sr_2IrO_4$ is one of the prime candidates for realizing exotic quantum spin orders owing to the subtle combination of spin-orbit coupling and electron correlation. Sensitive local magnetization measurement can serve as a powerful tool to study these kinds of systems with multiple competing spin orders since the comprehensive study of the spatially-varying magnetic responses provide crucial information of their energetics. Here, using sensitive magneto-optical Kerr effect measurements and spin Hamiltonian model calculations, we show that $Sr_2IrO_4$ has non-trivial domain structures which cannot be understood by conventional antiferromagnetism. This unconventional magnetic response exhibits broken symmetry along the Ir-O-Ir bond direction and is enhanced upon spin-flip transition or Rh-doping. Our analysis, based on possible stacking patterns of spins, shows that introduction of an additional rotational-symmetry breaking is essential to describe the magnetic behavior of $Sr_2Ir_{1-x}Rh_xO_4$, providing strong evidence for a nematic hidden order phase in this highly correlated spin-orbit Mott insulator.

## I. INTRODUCTION

The layered perovskite $Sr_2IrO_4$ has drawn significant attention due to its resemblance to cuprate superconductors in structural, electrical and magnetic properties [1–6]. Similar to the effect of doping in cuprates, the electrical and magnetic properties of $Sr_2IrO_4$ also change significantly upon doping, suggesting a potential to observe unconventional superconductivity in this system. Besides these similarities, $Sr_2IrO_4$ shows spin-orbit coupling (SOC) and electron correlation driven Mott insulating antiferromagnetism [7] and is a prominent candidate to realize an ideal spin-half Heisenberg model in a square lattice [1,6,8]. Particularly, slight Rh-doping is believed to decrease SOC, suppressing the antiferromagnetic (AFM) order and enlarging a "hidden" ordered (HO) phase above the Neel temperature ($T_N$) with an unknown order parameter whose coupling to the most of detection techniques are significantly weak, so deemed as hidden [9–12]. In the phase diagram, this HO phase resembles the pseudogap phase in cuprates, which surrounds the Mott insulating phase, raising questions about the nature of this phase and which kind of symmetry is broken. For undoped and slightly Rh-doped $Sr_2IrO_4$, second-harmonic generation (SHG), magnetic torque and polarized neutron diffraction studies proposed a loop-current order as an explanation of the HO phase [9–11], suggesting that exotic orbital effects may be reflected on the properties of the AFM phase due to strong SOC. More recently, the resonant X-ray study combined with Raman spectroscopy revealed the existence of a spin nematic order that develops above $T_N$ and coexists with the AFM phase in undoped $Sr_2IrO_4$ [13], providing further evidence that this spin-orbit coupled system is also highly correlated.

---

* These authors contributed equally to this work.

† Contact author: joonho.jang@snu.ac.kr

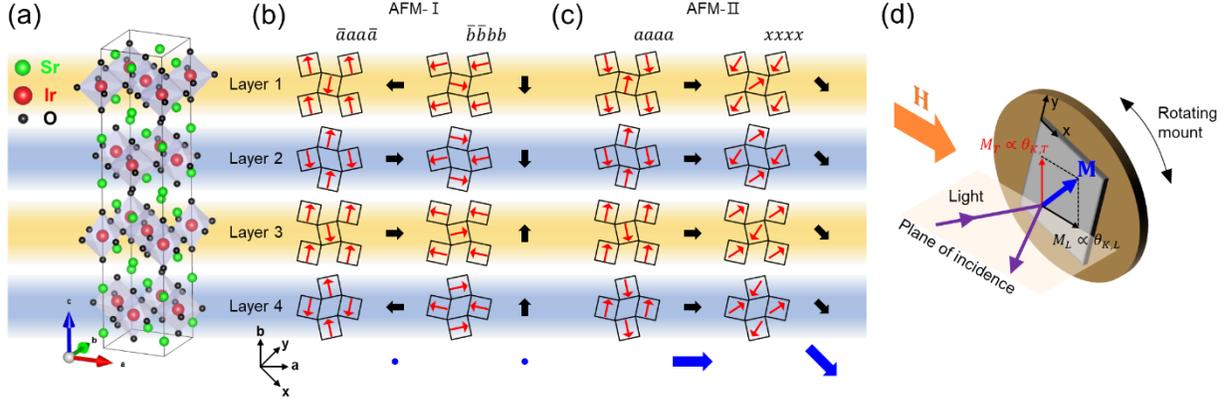

FIG. 1. Crystal structure and magnetic configuration of undoped $Sr_2IrO_4$. (a) The crystal structure of $Sr_2IrO_4$, which belongs to the tetragonal $I4_1/acd$ space group, is composed of canted $IrO_6$ octahedra [14,25]. The unit cell consists of four iridium-oxide layers. (b) (Top view) Ground state configurations in the AFM-I phase. (c) Potential magnetic configurations in the AFM-II phase for $H > H_C$. The thin red arrow indicates the pseudospin at each Ir site, the thick black arrow indicates the net moment of each layer, and the thick blue arrow indicates the net magnetization **M** in a unit cell. (d) Schematic of the MOKE measurement. $\theta_{K,L}$ ($\theta_{K,T}$) represents the longitudinal (transverse) Kerr angle, corresponding to the magnetization component parallel (perpendicular) to the external magnetic field **H**.

In order to understand the relationship between the HO phase and the AFM phase, an accurate characterization of the AFM phase is critical. Based on the concept of the spin-orbit coupled Mott insulator, the spin configurations of $Sr_2IrO_4$ are described by effective ½-spins ($J_{eff} = 1/2$) centered at Ir sites [14]. Notably, these pseudospins (red arrows in **Fig. 1(b)**) are confined to the basal plane of $IrO_2$, and the relative angles between pseudospins remain fixed regardless of the external magnetic field due to strong intralayer exchange interaction and the Dzyaloshinskii-Moriya interaction [15], which allows the net moment of each layer to be regarded as a single magnetic moment (black arrows in **Fig. 1(b)**) with a constant magnitude. The ground state configurations in zero magnetic field limit, however, lead to the cancellation of the total moment with layer moment configurations predominantly along the a-axis ($\bar{a}aa\bar{a}$) or b-axis ($\bar{b}\bar{b}bb$), depicted in **Fig. 1(b)** (AFM-I phase) [14,16]. Upon applying a magnetic field, owing to the relatively weak interlayer couplings, the moments of the individual layers rotate and can develop a net magnetization via the Zeeman interaction. Beyond the critical field $H_C$, all these moments abruptly flip to align in the same direction, resulting in weak ferromagnetic (WFM) behavior in $Sr_2IrO_4$, as shown in **Fig. 1(c)** (AFM-II phase) [14]. Various magnetic measurements have been conducted to investigate symmetry of the spin responses in the canted AFM phase [17–23], and significant deviations from the four-fold rotational symmetric magnetic responses anticipated from the tetragonal crystal structure are often observed [22–24]. The spin behaviors in this system are still yet to be fully understood but most previous experiments have relied on measurements of bulk properties, where inherent difficulties in resolving local orders and eliminating geometrical effects of a crystalline sample complicate the interpretation of the already complex quantum phenomena.

In this paper, we report scanning three-dimensional vector magneto-optical Kerr effect (MOKE) measurements of $Sr_2Ir_{1-x}Rh_xO_4$, that resolve the comprehensive local magnetic properties within individual domains. In undoped samples, the local responses exhibit peculiar two-fold rotational symmetry upon the spin-flip, with the direction of the symmetry axis mainly rotated 45° from the crystalline axes but varying spatially across a sample, suggesting a spontaneously broken rotational symmetry driven by an order coexisting with the conventional AFM phase. This observation is further corroborated with measurement of Rh-doped samples, where two types of domains with the respective easy-axes pointing in the directions whose angles are again 45° rotated from the crystalline axes, which is possibly related to the nematic HO phase.

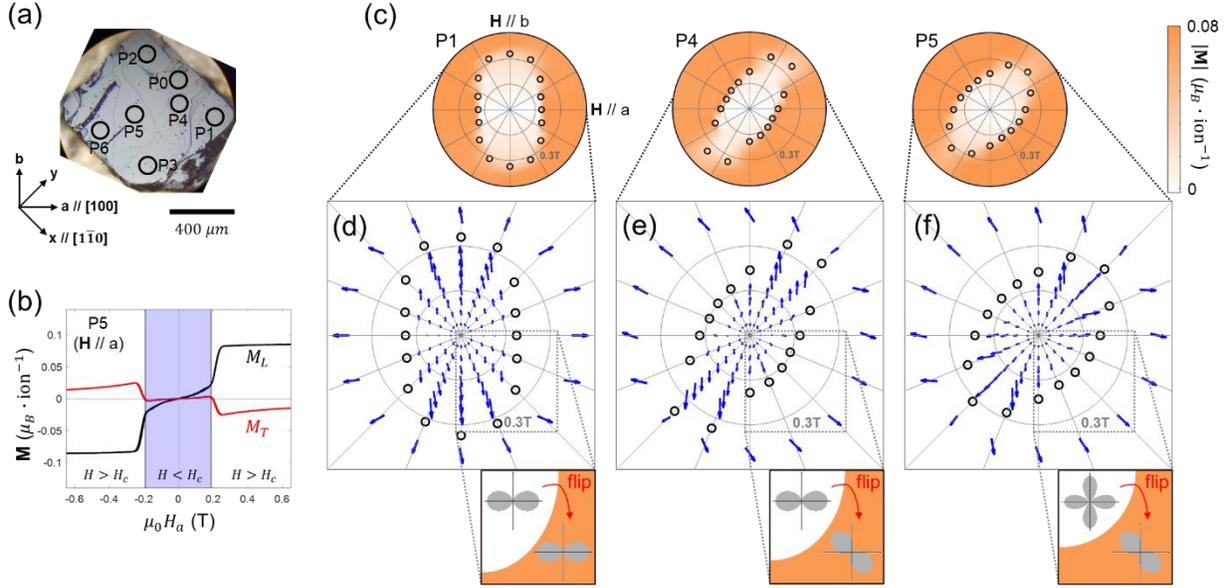

FIG. 2. Local magnetic behavior of undoped $Sr_2IrO_4$. (a) Optical microscope image of the sample. (b) Magnetization at position P5 in (a) as a function of the magnetic field applied along the a-axis. The shaded region represents $H < H_C$, and the unshaded region represents $H > H_C$. The Kerr data are converted to magnetization by comparing the saturated Kerr angle with the corresponding magnetization value of the reference [16]. (c) Magnetization magnitude |**M**| (color map) as a function of an in-plane magnetic field in a polar plot, at three different positions as indicated. Critical fields $H_C$ are indicated by black circles. (d)-(f) Detailed magnetic responses at (d) P1, (e) P4, and (f) P5. The blue arrows display net magnetizations **M** at various in-plane magnetic fields whose values are in the polar maps. For illustration, the arrows for $H < H_C$ are magnified four times, while for $H > H_C$, a single arrow is plotted per field direction. Insets in (d)-(f) depict the schematics of the uniaxial anisotropic energy terms extracted from the model calculation. All measurements were performed at $T = 5$ K.

## II. RESULTS

### A. MOKE measurement of undoped $Sr_2IrO_4$

The basal-plane rotational symmetry of the AFM phase in $Sr_2IrO_4$ is analyzed by measuring its local magnetization vector while sweeping the magnetic fields in-plane. To perform this experiment, we used a zero-area Sagnac interferometer to accurately determine the vector components of the local magnetization, by measuring both the longitudinal and transverse Kerr effects (**Fig. 1(d)** and **Appendix B**). As the magnetic field was always applied parallel to the plane of incidence, the longitudinal (transverse) Kerr signal represents the magnetization component parallel (perpendicular) to the magnetic field. With a sample mount rotating about the surface normal of the crystal, the relative direction of the in-plane magnetic field to the crystal axes of the sample could be controlled.

Initially, the magnetization vectors at selected positions (as denoted in **Fig. 2(a)**) were measured as a function of the magnetic field. **Figure 2(b)** shows the experimental results at position P5 when the magnetic field is applied along the a-axis.

Combining the longitudinal (black line) and transverse (red line) Kerr data, we can determine the full magnetization vector and represent its components in any crystallographic coordinate. Note that the jump in the magnetization at the critical field $\mu_0 H_C \sim 0.2$ T is the metamagnetic transition, in which the moments of the layers flip and point along the same direction [14]. Surprisingly, the transverse signal becomes non-zero beyond the transition, indicating that the direction of magnetization is not along the magnetic field even applied along the a-axis. This implies that the symmetry of the magnetic response is different from the symmetry given by the crystallographic axes anymore. To get a better picture of this phenomenon, we plotted the critical field $\mu_0 H_C$ as a function of magnetic field direction at the positions in **Fig. 2(c)**. The responses show two-fold rotational symmetry at all positions, but the symmetry direction, which is defined by the direction having the maximum value of $H_C$, is different for each position. Since the ground state spin configuration of $Sr_2IrO_4$ is known to be either $\bar{a}aa\bar{a}$ or $\bar{b}\bar{b}bb$, the extrema of $H_C$ are expected to be on either a- or b-axis but, at P4, P5, and P6, they are observed to be along the x- or y-axis, which is the Ir-O-Ir bond direction (see also **Supplemental Material, Fig. S1(a)** [26]).

For further analysis, the detailed view of the local magnetizations at positions P1, P4, and P5, which display representative behaviors, are shown in **Figs. 2(d)-2(f)**, respectively. In each panel, the blue arrows represent the magnetization vectors measured under the external magnetic fields whose coordinates are shown in a polar plot. In **Fig. 2(d)**, we show the data at position P1, which is one of the most conventional and thus the simplest to interpret. The magnetization vectors for $H < H_C$ lie noticeably parallel to the b-axis throughout the directions of the magnetic field until the spin-flip transition occurs. When a small magnetic field is applied in the AFM-I phase, each layer moment has no option but tilts slightly. As a result, any net magnetization would be perpendicular to the orientation of the layer moments of the ground state. Based on this argument, the observed net magnetization along the b-axis should be ascribed to the magnetic response of a region with $\bar{a}aa\bar{a}$ (left panel in **Fig. 1(b)**) – or its Kramers pair $a\bar{a}\bar{a}a$ – and rules out other potential ground state spin configurations – such as $\bar{b}\bar{b}bb$ – where domains that break the symmetry of $\bar{a}aa\bar{a}$ and $\bar{b}\bar{b}bb$ configurations were reported and attributed to crystal twin boundaries [24,27,28]. Beyond the spin-flip ($H > H_C$), all the layer moments point to nearly the same directions due to the Zeeman energy and thus follow the magnetic fields more freely, but there still exists preference to align along the a-axis. In our model Hamiltonian analysis, the responses at P1 below and above the spin-flip are well explained by the single Hamiltonian term $K_{ab}$ (**Appendix C**). Qualitatively similarly, the magnetic behaviors at positions P2 and P3 (see **Supplemental Material, Figs. S1(b)** and **S1(c)** [26]) show nearly conventional behaviors of $\bar{b}\bar{b}bb$ and $\bar{a}aa\bar{a}$, respectively, but with anomalous distortion in the shape of $H_C$.

Interestingly, the effect of distortion becomes stronger at position P4. In **Fig. 2(e)**, the magnetization below the spin-flip transition is along the b-axis also similar to the case of position P1, which infers that the ground state configuration is along the a-axis. However, after the spin-flip transition, the magnetization prefers to align along the x-axis, unexpected from the given crystalline symmetry. Without any *a priori* reason for spins to tend to align along the x- or y-axis, this seems to suggest spontaneously-broken rotational symmetry and the existence of a nematic order that couples with the WFM moment and provides uniaxial preference to the Ir-O-Ir bond direction. Further corroborating the effect, the data at position P5 and P6 are plotted in **Fig. 2(f)** and **Fig. S1(d)** [26], respectively. First, the magnetic behaviors for $H < H_C$ are those expected from a single domain with the ground state configurations $\bar{b}\bar{b}bb$ and $\bar{a}aa\bar{a}$ nearly degenerate and coexisting. Nevertheless, when $H > H_C$, the effect of coupling with the bond-directional order discussed at the position P4 is again observed, evidencing that this order represents an independent mechanism distinct from the effect that locks the moments to one of the tetragonal crystal axes.

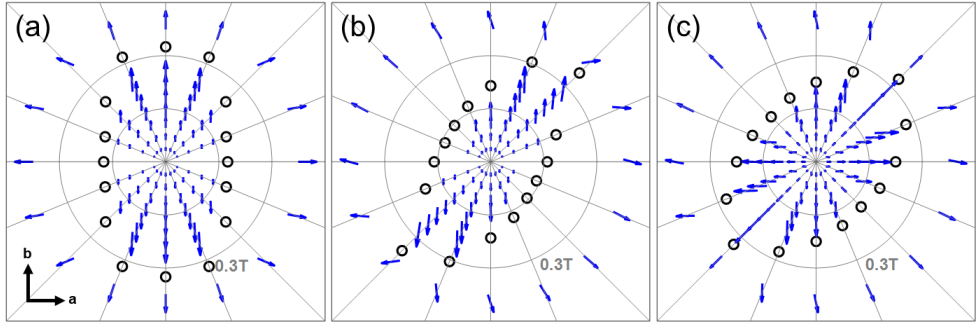

FIG. 3. Calculated magnetic responses at different positions in the undoped sample. (a)-(c) Model calculations of magnetic responses at (a) P1, (b) P4, and (c) P5. Critical fields $H_C$ are indicated by black circles, and the net magnetizations **M** are indicated by blue arrows. Note that the size of the arrows for $H < H_C$ is magnified four times for illustration. The parameters used in the model calculation are provided in **Supplemental Material, Sec. S2** [26].

The bond-directional order can be associated with an energy term that gives preference along the x- or y-axis and couples strongly with moments after the spin-flip transition. To implement phenomenologically these observations in the model calculations (**Appendix C**), we introduce the following additional Hamiltonian term:

$$H_{xy} = -N_{xy}\left(\left(\frac{1}{4}\sum_{i=1}^{4}cos\alpha_i\right)^2 - \left(\frac{1}{4}\sum_{i=1}^{4}sin\alpha_i\right)^2\right), (1)$$

where $\alpha_i$ refers to the angle of the layer moment of the $i$-th layer in a unit cell with respect to the x-axis. Since $H_{xy}$ is proportional to $-N_{xy}(M_x^2 - M_y^2)$ where $M_x$ ($M_y$) is the magnetization component along the x- (y-) axis, it has little effect prior to the spin-flip transition. After the transition, $H_{xy}$ acts to align the net magnetization along the x- or y-axis. The Hamiltonian $H_{xy}$ can be interpreted as a result of the exchange interaction between the pseudospins and the background electronic subsystem with anisotropic susceptibility along the Ir-O-Ir bond direction (**Supplemental Material, Sec. S3** [26]). By introducing $H_{xy}$, we quantitatively reproduce the critical features of the data [**Figs. 3(b)** and **3(c)**], to the level we are strongly convinced that the correct model Hamiltonian with minimal terms is used. Thus, this model supports the picture that the WFM moment in the AFM-II phase, which suddenly develops above the spin-flip transition, triggers strong interaction between the Ir spins and the anisotropic nematic order, resulting in the emergent two-fold symmetry selectively observed in the AFM-II phase. We note that introducing a finite value of the pseudo-Jahn-Teller (pseudo-JT) effect is also required to describe the sudden jump of the magnetization [29]. By comparing the experimental results with the model calculations [**Figs. 2(d)-2(f)** and **Figs. 3(a)-3(c)**], we determined that the $xy$ type orthorhombic distortion (denoted as $\Gamma_1$ in **Appendix C**) is dominant in the undoped sample [16,29].

As discussed, the spin-flip transition represents a point that separates the low-field limit where the a-b symmetry breaking effect is dominant and the high-field limit where the moments strongly couple to the bond-directional nematic order. To study the spatial distribution of these effects, we present scanning MOKE images of the undoped sample in **Fig. 4**. First of all, we measured remnant magnetization in the absence of an external magnetic field to distinguish non-ideal regions, as shown in **Fig. 4(a)**; the sample shows zero magnetization as expected for the AFM phase except for a small region with finite remnant magnetization which is coincident with the position P0 in **Fig. 2(a)**. This region displays a WFM response along the x-axis with $T_N$ of ~200 K (**Supplemental Material, Figs. S3** and **S4** [26]), similar to that of a weakly doped $Sr_2IrO_4$, and we attribute it to originate from the oxygen deficiency [30]. Increasing the magnetic field, the sample shows a unique spatial pattern of magnetization as shown in **Fig. 4(b)**. Then, to investigate the spatial distribution of the distinct broken rotational symmetries, such as a-, b-, x- and y-axis preferences discussed in **Fig. 2**, we devise an efficient measure for visualization with a differential susceptibility tensor $\chi_{ij} = \frac{\partial M_i}{\partial H_j}|_{H_0}$ ( $i,j = a,b$ ), where the quantity is evaluated at $H = H_0$. For $H < H_C$, where the phases preferring the a-axis ($\bar{a}aa\bar{a}$) and the b-axis ($\bar{b}\bar{b}bb$) are to be distinguished, $\chi_{aa}$ evaluated at $H_0 = 0$ is plotted in **Fig. 4(c)**. The region of non-zero values is where $\bar{b}\bar{b}bb$ configuration is present while the region of near zero values means that the area allows $\bar{a}aa\bar{a}$ as the only allowed configuration. For $H > H_C$, due to the emergence of coupling to a bond-directional order, the symmetry breakings along a-, b-, x- and y-axes are to be distinguished. We use $\chi_{ba}$ evaluated at $H_0 > H_C$ in **Fig. 4(d)**, where the positive (negative) value indicates that the nematic order is along the x-axis (y-axis). Note that if a phase has symmetry breaking along the a- or b-axis with weak bond-directional order, $\chi_{ba}$ would be zero, which is visible as yellow-colored regions in **Fig. 4(d)** (see also **Supplemental Material, Sec. S6** [26]). Remarkably, a comparison of two images indicates that they both have domain structures. The a-b symmetry breaking in $H < H_C$ is known to be mostly dictated by crystal twinning [24,27,28]; on the other hand, the image for x-y symmetry breaking at $H > H_C$ apparently shows a different spatial pattern across the sample (see also **Supplemental Material, Fig. S5** [26]), supporting that a unique mechanism plays a role after the spin-flip transition. The observation of the macroscopic-sized regions with a sudden and distinctive change of rotational symmetry to the x- or y-axis above $H_C$ evidences the formation of domains of a bond-directional order, which cannot be accounted for by any conventional AFM picture of $Sr_2IrO_4$.

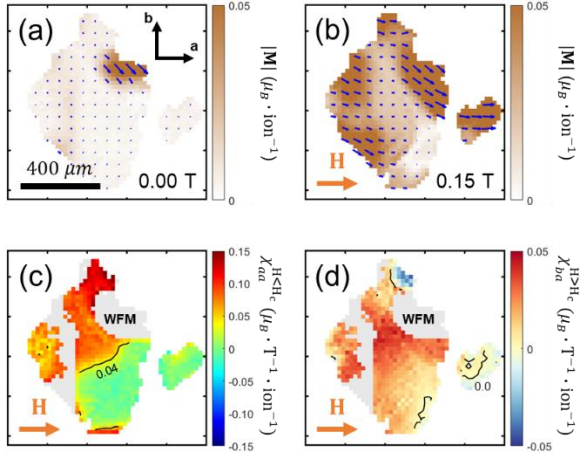

FIG. 4. Scanning MOKE image of undoped $Sr_2IrO_4$. (a),(b) Magnetization image of the sample at (a) $\mu_0H = 0$ T and (b) $\mu_0H = 0.15$ T. Arrows indicate the direction and magnitude of the local magnetization **M**. (c),(d) Differential magnetic susceptibility images. (c) $\chi_{aa}$ for $H < H_C$ and (d) $\chi_{ba}$ for $H > H_C$. The magnetic field is applied along the a-axis. Grey areas indicate WFM regions in the absence of an external magnetic field.

### B. MOKE measurement of Rh-doped $Sr_2IrO_4$ ($Sr_2Ir_{1-x}Rh_xO_4$)

We have established that the coupling to the bond-directional nematic order appears only in the AFM-II phase of undoped $Sr_2IrO_4$, after the spin-flip transition. The AFM-II phase is also known to be induced by slight Rh-doping [31–33], and we thus measured Rh-doped $Sr_2IrO_4$ to determine whether the spin-flipped state shows the same bond-directional magnetic behavior.

To identify the rotational symmetry of Rh-doped $Sr_2IrO_4$, we performed the rotating-field MOKE measurements at a representative position in a 3% Rh-doped sample. As shown in **Fig. 5(a)**, the magnitude remained to be nearly constant for a wide range of magnetic field strength and angles, indicating a saturation of moments; this confirms that all the moments of layers are already flipped and point to the same direction, which is the characteristic of the AFM-II phase. More importantly, it shows two-fold rotational symmetric magnetic responses with the symmetry axis along the x-axis, evidencing that the spin-flipped state exhibits the same bond-directional nematic behavior as undoped $Sr_2IrO_4$. In **Fig. 5(b)**, we plotted $M_T/M_L$ (a quantity to represent the misalignment of the magnetization vector with respect to the magnetic field direction) as a function of $\phi$, and compared this with the calculation based on the same model Hamiltonian used in the undoped case. The calculation reproduces very accurately all features of the experimental data, showing the easy axis 45° rotated from the tetragonal crystal axes. It becomes apparent that the bond-directional order influences dominantly in the AFM-II phases of all samples, regardless of doping level (**Supplemental Material, Figs. S3** and **S6** [26]).

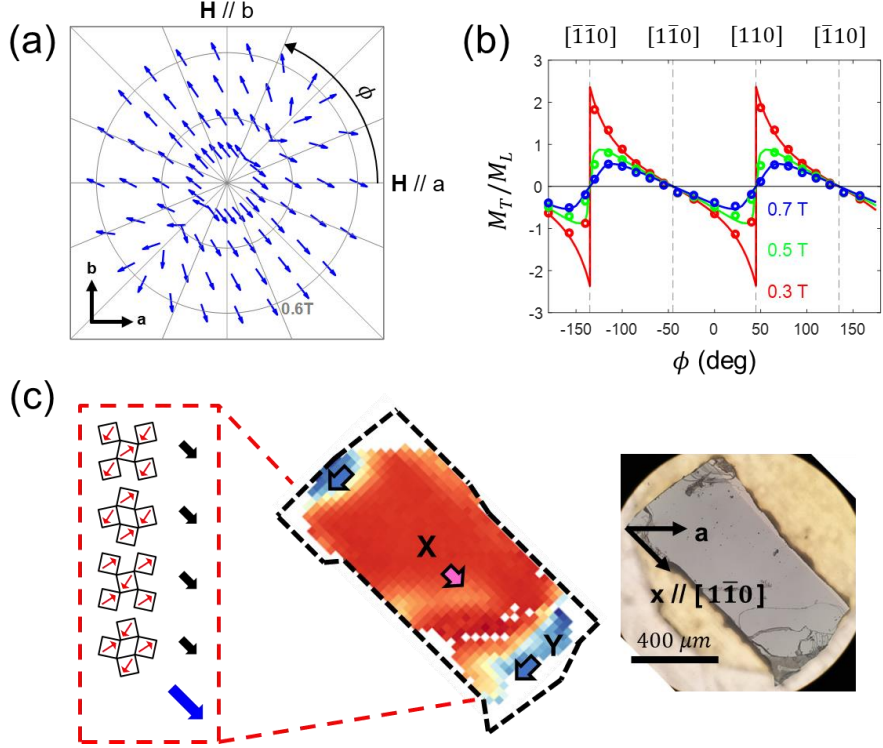

FIG. 5. MOKE measurement and model calculation of the 3% Rh-doped $Sr_2IrO_4$. (a) Quiver plot of local magnetic responses, where blue arrows indicate the magnetization vectors **M**. (b) Comparison between experimental data (circles) and model calculation (lines). $\phi$ is the angle of the magnetic field vector with respect to the a-axis. In the model calculation, we used parameters of $J_{1c}$ = -13.0 μeV, $J_{2c}$ = 3.0 μeV, $N_{xy}$ = 1.9 μeV, $\Gamma_2$ = 1.2 μeV, and an effective parameter $E_{Th}$ = 0.4 μeV, which was introduced to describe magnetic hysteresis (see model Hamiltonian in **Appendix C** for details of each parameter). (c) Magnetic configuration of domain X and scanning MOKE image at zero magnetic field, showing magnetizations aligned along the x-axis (denoted as X) or y-axis (denoted as Y). All measurements were performed at $T$ = 5 K.

The most critical feature is that the magnetization behaves similar to a ferromagnet with an easy axis along the x-axis. This behavior is captured in the model calculation, which we attribute to Rh-doping-induced sign-change in interlayer interactions [32] and the bond-directional order with positive $N_{xy}$ (see Eq.1). As shown in **Fig. 5(c)**, the positive $N_{xy}$ term enforces the ground state configuration to be $xxxx$, excluding any possibility of a domain mixed of conventionally-argued $aaaa$ and $\bar{b}\bar{b}\bar{b}\bar{b}$. It is worthy to note that we also had to introduce the $x^2$- $y^2$ type orthorhombic distortion (denoted as $\Gamma_2$ in **Appendix C**) to closely reproduce the kink indicated by a green arrow in **Fig. S7** [26]. This term confirms the previous conjecture that the ground state magnetic ordering affects the nature of the pseudo-JT effect [29]; i.e., the $\bar{a}aa\bar{a}$ or $\bar{b}\bar{b}bb$ state in undoped $Sr_2IrO_4$ is related with lattice distortion along the a- and b-axes ($xy$ type), whereas the $xxxx$ or $yyyy$ state in Rh-doped $Sr_2IrO_4$ is associated with lattice distortion along the x- and y-axes ($x^2$-$y^2$ type).

To study the spatial distribution of the nematic order, zero-magnetic-field scanning MOKE images of Rh-doped samples of 3% and 5% doping are shown in **Fig. 5(c)** and **Fig. S6(a)** [26], respectively. Indeed, the magnetizations are distributed exclusively along the x- or y-axis, and the images display distinctive domain structures with two mutually orthogonal axes (denoted as X and Y). In the model calculation, the magnetic field response of any region can be modeled using the $N_{xy}$ term, simply by changing its sign (**Supplemental Material, Figs. S6(b) - S6(i)** [26]). To our surprise, the observation of X and Y domains contrasts with previous works based on Bragg peaks [31,32], which suggested that magnetization lies along the a- or b-axis in doped samples. It should be noted, however, that the previous measurements provide information averaged over the substantial area that represents properties of a whole sample, which importantly differ from our study that directly measures local characteristics, in the presence of the spatially-varying bond-directional order.

### III. DISCUSSION AND SUMMARY

This work represents systematic local profiling of the full 3D magnetization of spin orders, measured across the whole surface of high-quality single crystals $Sr_2Ir_{1-x}Rh_xO_4$. The data acquired by sweeping in-plane magnetic fields provide a comprehensive picture of the energetics of spins in this canted AFM phase and reveal the previously unknown features and hidden broken-symmetries, especially through the response above the spin-flip transition.

While our analysis provides a strong case for electron nematicity in this system, the microscopic mechanism that could drive the nematicity is still unspecified. The current work however places stringent constraints on theoretical models that attempt to implement the nematicity. To our knowledge, there are two major proposals in the literature for a nematic order that can be present in this system; a loop-current order [9–11] and a spin nematic order [13]. The existence of the bond-directional loop-current order, which preserves translational symmetry but is expected to break inversion, time-reversal and rotational symmetry, could give rise to nematicity of the coexisting electron system but is only partially consistent with our observations; spontaneous breaking of time-reversal symmetry is not observed in any of samples with our sensitive Sagnac interferometer above $T_N$ (Ref. [13]; **Supplemental Material, Fig. S4** [26]). On the other hand, the recently discovered spin nematic phase preserves time-reversal symmetry. The quadrupolar order of the spin nematicity and the dipolar order of the canted AFM phase mutually constrain their directions with each other [13] and thus, in our measurement in $H > H_C$, we effectively rotate the ferromagnetic component of the AFM phase and the ferro-quadrupolar component of spin nematicity together by rotating the external magnetic field. Since the quadrupole is known to prefer to point to the bond directions [13,34,35], which is intuitively understood based on the fact that a quadrupole of the spin nematic phase is formed by two spins at the nearest neighbor sites sharing the Ir-O-Ir bond, the energetic preference of the mutually-constraint dipole to align to one of the bond directions is naturally explained. The bond-directional response of the canted AFM dipoles, combined with the observed preservation of time-reversal symmetry, thus provides evidence of the spin nematic phase as the mechanism behind the observed nematicity.

We also mention that this spatially distributed domain structure cannot be explained by sample strains [36] or by a peculiar surface magnetization suggested in the SHG experiment [24]. First, there is no reason to believe that directions of strain from extrinsic sources (i.e., different thermal contractions between the sample and grease) would align along particular high-symmetric crystal axes, such as a-, b-, x- and y-axis, and be distributed over the length scale of 100 μm and 1000 μm, which is the typical size of a domain observed (**Fig. 5(c)** and **Fig. S6(a)** [26]). Second, there is no indication of suppression of $T_N$ in any of the crystals we measured. In addition, our optical measurement accesses at least hundreds of unit cells from the surface of a crystal, estimated based on the optical penetration depth, which excludes any possibility of a signal originating from a certain surface termination or reconstruction. We also note a recent study that suggested Ir-O-Ir bond-directional spin orientation in insulating $SrIrO_3$ monolayer film while the microscopic mechanism is not proposed [37].

## ACKNOWLEDGMENTS

This work was supported by the National Research Foundation of Korea grants funded by the Ministry of Science and ICT (Grant Nos. 2019R1C1C1006520, RS-2020-NR049536, and RS-2023-00258359), the Institute for Basic Science of Korea (Grant No. IBS-R009- D1), SNU Core Center for Physical Property Measurements at Extreme Physical Conditions (Grant No. 2021R1A6C101B418), and Creative-Pioneering Researcher Program through Seoul National University and Samsung Electronics. H.H. was supported by the Basic Science Research Program through the National Research Foundation of Korea (NRF) funded by the Ministry of Education (Grant No. RS-2024-00409528). B.J.K was supported by the Samsung Science and Technology Foundation (Project SSTF-BA2201-04) and National Research Foundation of Korea (Grant RS-2024-00360303) funded by the Korean Government (Ministry of Science and ICT).

## Appendices
### APPENDIX A: Sample preparation

The $Sr_2Ir_{1-x}Rh_xO_4$ single crystals were grown using the flux method. Stoichiometric quantities of $SrCO_3$, $IrO_2$, and $Rh_2O_3$ powders were thoroughly mixed with $SrCl_2$ flux. The powder-to-flux molar ratio was 1:7, based on previously reported synthesis methods [30,38,39]. The mixed materials were placed in a platinum crucible. The growth procedure involved heating the mixture to 1300 °C and then slowly cooling it to 800 °C at a rate of 2 °C per hour, followed by furnace cooling. The doping concentration of each sample was confirmed using an energy-dispersive X-ray spectroscopy machine (Hitachi S-3400N). The measured doping concentrations for the x = 0.03 and x = 0.05 samples were 0.0369 and 0.0550, respectively, showing no significant deviation from the nominal ones. Neel temperature measured by MOKE further confirms the doping level (**Supplemental Material, Fig. S4** [26]).

### APPENDIX B: Magneto-optical Kerr effect (MOKE) measurement

We use the normal-incidence (oblique-incidence) zero-area Sagnac interferometer operating at 1550 nm wavelength to measure local out-of-plane (in-plane) magnetic properties [40,41]. This interferometry-based scheme provides superior stability and reliability for detecting magnetic signals from a sample, as it rejects all reciprocal effects, such as linear birefringence and optical activity, in both the optics and the sample. For the normal-incidence MOKE measurement, we didn't find any discernible polar Kerr signal except for WFM domain boundaries or near the sample edges under the noise limit, confirming spins lying on the $IrO_2$ plane. For the oblique-incidence MOKE measurement, a beam of light is incident on a sample with an angle of $\theta_{inc}$~15°. As shown in **Fig. 1(d)**, magnitude |**M**| (Eq. 2) and direction Ω

(Eq. 3) of the local magnetization are calculated by measuring both longitudinal Kerr angle $\theta_{K,L} = Re(\frac{2sin\theta_{inc}Q}{\varepsilon_s-1})M_L$ and transverse Kerr angle $\theta_{K,T} = Re(\frac{2sin\theta_{inc}Q}{\varepsilon_s-1})M_T$ [40]:

$$|\mathbf{M}| = \sqrt{M_L^2 + M_T^2} \propto \sqrt{\theta_{K,L}^2 + \theta_{K,T}^2}, \quad (2)$$

$$\Omega = tan^{-1}(\theta_{K,T}/\theta_{K,L}), \quad (3)$$

where Voigt parameter $Q$ and dielectric constant $\varepsilon_s$ at wavelengths of 1550 nm are material dependent parameters. In the current study, we apply the in-plane magnetic field along the plane-of-incidence; thus, $M_L$ ($M_T$) represents the in-plane magnetization component of a sample that is parallel (perpendicular) to the plane-of-incidence and the magnetic field. Then we rotate the sample, to control the angle between the magnetic field and the crystallographic axes of the sample. A beam spot is approximately 50 μm in diameter. We confirm that the optical heating is smaller than 1 K as the incident optical power is maintained below 1 mW. For all data presented, we subtract temperature independent background signal from optical windows which is linearly proportional to the strength of the external magnetic field (**Supplemental Material, Figs. S8** and **S9** [26]).

### APPENDIX C: Model Hamiltonian

The Hamiltonian for the first- and second-nearest-interlayer interactions is written in the form [16]:

$$H_{inter} = \frac{1}{N}\sum_{<i\,j>} J_{1c}S_i \cdot S_j \pm \Delta_c(S_i^a S_j^a - S_i^b S_j^b) \\ + \frac{1}{N}\sum_{<i\,j>} J_{2c}S_i \cdot S_j, \quad (4)$$

where the first (second) $<i\,j>$ refers to the nearest pseudospin pairs in the first (second) adjacent layers, $N$ refers to the number of Ir-atoms, and $a$ and $b$ denote the directions along the crystal axes.

Note that one pseudospin $S_i$ has eight (two) nearest pseudospins $S_j$ in the first (second) two nearest layers. $J_{1c}$ and $J_{2c}$ determine the exchange energy of the corresponding pseudospin pairs. $\Delta_c$ accounts for orbital characteristics of pseudospins and gives the direction dependence in the nearest interlayer coupling [16,42].

Additionally, the strong spin-orbit coupling in $Sr_2IrO_4$ allows the pseudospins to be coupled to lattice vibrations, which is called pseudo-JT effect. The Hamiltonian of the pseudo-JT effect is written in the form [16,29]:

$$H_{JT} = -\Gamma_1\left(\frac{1}{N}\sum_{<i\,j>} S_i^a S_j^a - S_i^b S_j^b\right)^2 \\ -\Gamma_2\left(\frac{1}{N}\sum_{<i\,j>} S_i^a S_j^b + S_i^b S_j^a\right)^2, \quad (5)$$

where $<i\,j>$ refers to the nearest intralayer pseudospin pairs. Note that $\Gamma_1$ and $\Gamma_2$ define the preference for $xy$ and $x^2$-$y^2$ type orthorhombic distortion of $IrO_6$ octahedra. The observation of metamagnetic transition in $Sr_2IrO_4$ can be explained with the pseudo-JT effect [29].

Using the Hamiltonian $H = H_{inter} + H_{JT}$ from Eqs. 4, 5, the magnetic configurations of $\bar{a}aa\bar{a}$ and $\bar{b}\bar{b}bb$, shown in **Fig. 1(b)**, and their Kramers pairs are the most stable [14,16] (**Supplemental Material, Sec. S10** [26]). This description could support the observations of magnetic Bragg peaks implying the magnetic configurations [14,43]. However, there are reports that moments didn't have both $\bar{a}aa\bar{a}$ and $\bar{b}\bar{b}bb$ configurations as the ground state, but only one of them as the ground state [24,27,28], which cannot be explained by the four-fold Hamiltonian. To explain this, it is necessary to introduce a Hamiltonian that breaks the four-fold rotational symmetry and locks the moments to a unique in-plane crystal axis, which can be written in the form:

$$H_{aniso} = -\frac{1}{N}\sum_{<i j>} K_{ab}(S_i^a S_j^a - S_i^b S_j^b), \quad (6)$$

where $<i\ j>$ refers to the nearest intralayer pseudospin pairs.

We also introduced a Hamiltonian $H_{xy}$ to describe the new bond-directional order in Eq. 1. Based on these arguments, we calculated the energy $E$ derived from the Hamiltonian $H = H_{inter} + H_{JT} + H_{aniso} + H_{xy}$ with Zeeman energy in Eq. 7.

$$\begin{aligned} E &= \frac{j_{1c}}{4}\sum_{i=1}^{4} cos(\alpha_i - \alpha_{i+1}) + \frac{j_{2c}}{4}\sum_{i=1}^{4} cos(\alpha_i - \alpha_{i+2}) \\ &+ \frac{\delta_c}{4}\sum_{i=1}^{4}(-1)^{i+1} sin(\alpha_i + \alpha_{i+1}) \\ &- \gamma_1\left(\frac{1}{4}\sum_{i=1}^{4} sin(2\alpha_i)\right)^2 - \gamma_2\left(\frac{1}{4}\sum_{i=1}^{4} cos(2\alpha_i)\right)^2 \\ &- \frac{k_{ab}}{4}\sum_{i=1}^{4} sin(2\alpha_i) \\ &- N_{xy}\left\{\left(\frac{1}{4}\sum_{i=1}^{4} cos\alpha_i\right)^2 - \left(\frac{1}{4}\sum_{i=1}^{4} sin\alpha_i\right)^2\right\} \\ &- \frac{h}{4}\sum_{i=1}^{4} cos\left(\alpha_i - \phi - \frac{\pi}{4}\right), \quad (7) \end{aligned}$$

where $\alpha_1\ (=\alpha_5)$, $\alpha_2\ (=\alpha_6)$, $\alpha_3$, $\alpha_4$ are the angles of the layer moments with respect to the x-axis. $\phi$ is the angle of the external magnetic field with respect to the a-axis. Note that $j_{1c} = 4S^2 J_{1c} sin^2\varphi$, $j_{2c} = -S^2 J_{2c} cos2\varphi$, $\delta_c = 4S^2\Delta_c cos^2\varphi$, $\gamma_1 = 4S^4\Gamma_1$, $\gamma_2 = 4S^4\Gamma_2$, $k_{ab} = 2S^2 K_{ab}$, $h = g\mu_0\mu_B SH sin\varphi$, with $S = 1/2$, $\varphi = 12°$, $g = 2$, and $\mu_B = 57.88$ μeV/T. For more details on the model calculation, see references [16,29]. To induce hysteric behavior in the model calculation, we introduce a constant decision threshold $E_{Th}$. Similar to the rotating-field MOKE measurement, the model calculation is performed by sweeping the magnetic field from positive to negative and then from negative to positive. For the initial magnetic structure, we determine the spin configurations that minimize the energy of the four spin layers, whose energy $E_G$ represents the global minimum energy. For subsequent magnetic structures, we compare $E_G$ with the local minimum energy $E_L$. $E_L$ is determined using the Nelder-Mead minimization method from the previous spin configuration. If $E_L - E_G < E_{Th}$ ($E_L - E_G > E_{Th}$), the spin configuration whose energy is the local (global) minimum is chosen for the next spin configuration.

[1] J. Bertinshaw, Y. K. Kim, G. Khaliullin, and B. J. Kim, *Square Lattice Iridates*, Annu. Rev. Condens. Matter Phys. **10**, 315 (2019).

[2] C. Lu and J. Liu, *The $J_{eff}$ = 1/2 Antiferromagnet $Sr_2IrO_4$: A Golden Avenue toward New Physics and Functions*, Advanced Materials **32**, 1904508 (2020).

[3] Y. Cao et al., *Hallmarks of the Mott-metal crossover in the hole-doped pseudospin-1/2 Mott insulator $Sr_2IrO_4$*, Nat. Commun. **7**, 11367 (2016).

[4] Y. K. Kim, N. H. Sung, J. D. Denlinger, and B. J. Kim, *Observation of a d-wave gap in electron-doped $Sr_2IrO_4$*, Nat. Phys. **12**, 37 (2016).

[5] Y. K. Kim, O. Krupin, J. D. Denlinger, A. Bostwick, E. Rotenberg, Q. Zhao, J. F. Mitchell, J. W. Allen, and B. J. Kim, *Fermi arcs in a doped pseudospin-1/2 Heisenberg antiferromagnet*, Science **345**, 187 (2014).

[6] J. Kim et al., *Magnetic Excitation Spectra of $Sr_2IrO_4$ Probed by Resonant Inelastic X-Ray Scattering: Establishing Links to Cuprate Superconductors*, Phys. Rev. Lett. **108**, 177003 (2012).

[7] B. J. Kim et al., *Novel $J_{eff}$ = 1/2 Mott State Induced by Relativistic Spin-Orbit Coupling in $Sr_2IrO_4$*, Phys. Rev. Lett. **101**, 076402 (2008).

[8] J. G. Rau, E. K.-H. Lee, and H.-Y. Kee, *Spin-Orbit Physics Giving Rise to Novel Phases in Correlated Systems: Iridates and Related*

*Materials*, Annu. Rev. Condens. Matter Phys. **7**, 195 (2016).

[9] L. Zhao, D. H. Torchinsky, H. Chu, V. Ivanov, R. Lifshitz, R. Flint, T. Qi, G. Cao, and D. Hsieh, *Evidence of an odd-parity hidden order in a spin–orbit coupled correlated iridate*, Nat. Phys. **12**, 32 (2016).

[10] J. Jeong, Y. Sidis, A. Louat, V. Brouet, and P. Bourges, *Time-reversal symmetry breaking hidden order in* $Sr_2(Ir,Rh)O_4$, Nat. Commun. **8**, 15119 (2017).

[11] H. Murayama et al., *Bond Directional Anapole Order in a Spin-Orbit Coupled Mott Insulator* $Sr_2(Ir_{1-x}Rh_x)O_4$, Phys. Rev. X **11**, 011021 (2021).

[12] C. Tan et al., *Slow magnetic fluctuations and critical slowing down in* $Sr_2Ir_{1-x}Rh_xO_4$, Phys. Rev. B **101**, 195108 (2020).

[13] H. Kim et al., *Quantum spin nematic phase in a square-lattice iridate*, Nature **625**, 264 (2024).

[14] B. J. Kim, H. Ohsumi, T. Komesu, S. Sakai, T. Morita, H. Takagi, and T. Arima, *Phase-Sensitive Observation of a Spin-Orbital Mott State in* $Sr_2IrO_4$, Science **323**, 1329 (2009).

[15] G. Jackeli and G. Khaliullin, *Mott Insulators in the Strong Spin-Orbit Coupling Limit: From Heisenberg to a Quantum Compass and Kitaev Models*, Phys. Rev. Lett. **102**, 017205 (2009).

[16] J. Porras et al., *Pseudospin-lattice coupling in the spin-orbit Mott insulator* $Sr_2IrO_4$, Phys. Rev. B **99**, 085125 (2019).

[17] T. Takayama, A. Matsumoto, G. Jackeli, and H. Takagi, *Model analysis of magnetic susceptibility of* $Sr_2IrO_4$: *A two-dimensional* $J_{eff} = 1/2$ *Heisenberg system with competing interlayer couplings*, Phys. Rev. B **94**, 224420 (2016).

[18] L. Fruchter, D. Colson, and V. Brouet, *Magnetic critical properties and basal-plane anisotropy of* $Sr_2IrO_4$, J. Phys.: Condens. Matter **28**, 126003 (2016).

[19] M. Nauman, Y. Hong, T. Hussain, M. S. Seo, S. Y. Park, N. Lee, Y. J. Choi, W. Kang, and Y. Jo, *In-plane magnetic anisotropy in strontium iridate* $Sr_2IrO_4$, Phys. Rev. B **96**, 155102 (2017).

[20] M. Nauman, T. Hussain, J. Choi, N. Lee, Y. J. Choi, W. Kang, and Y. Jo, *Low-field magnetic anisotropy of* $Sr_2IrO_4$, J. Phys.: Condens. Matter **34**, 135802 (2022).

[21] H. Wang, W. Wang, N. Hu, T. Duan, S. Yuan, S. Dong, C. Lu, and J.-M. Liu, *Persistent Large Anisotropic Magnetoresistance and Insulator-to-Metal Transition in Spin-Orbit-Coupled* $Sr_2(Ir_{1-x}Ga_x)O_4$ *for Antiferromagnetic Spintronics*, Phys. Rev. Applied **10**, 014025 (2018).

[22] H. Wang, C. Lu, J. Chen, Y. Liu, S. L. Yuan, S.-W. Cheong, S. Dong, and J.-M. Liu, *Giant anisotropic magnetoresistance and nonvolatile memory in canted antiferromagnet* $Sr_2IrO_4$, Nat. Commun. **10**, 2280 (2019).

[23] C. Wang, H. Seinige, G. Cao, J.-S. Zhou, J. B. Goodenough, and M. Tsoi, *Anisotropic Magnetoresistance in Antiferromagnetic* $Sr_2IrO_4$, Phys. Rev. X **4**, 041034 (2014).

[24] K. L. Seyler, A. De La Torre, Z. Porter, E. Zoghlin, R. Polski, M. Nguyen, S. Nadj-Perge, S. D. Wilson, and D. Hsieh, *Spin-orbit-enhanced magnetic surface second-harmonic generation in* $Sr_2IrO_4$, Phys. Rev. B **102**, 201113 (2020).

[25] J. Kim, H. Kim, H.-W. J. Kim, S. Park, J.-K. Kim, J. Kwon, J. Kim, H. W. Seo, J. S. Kim, and B. J. Kim, *Single crystal growth of iridates without platinum impurities*, Phys. Rev. Materials **6**, 103401 (2022).

[26] See Supplemental Material for supporting data and detailed explanation on the model calculations.

[27] F. Ye, S. Chi, B. C. Chakoumakos, J. A. Fernandez-Baca, T. Qi, and G. Cao, *Magnetic and crystal structures of* $Sr_2IrO_4$: *A neutron diffraction study*, Phys. Rev. B **87**,

140406 (2013).

[28] C. Dhital, T. Hogan, Z. Yamani, C. De La Cruz, X. Chen, S. Khadka, Z. Ren, and S. D. Wilson, *Neutron scattering study of correlated phase behavior in* $Sr_2IrO_4$, Phys. Rev. B **87**, 144405 (2013).

[29] H. Liu and G. Khaliullin, *Pseudo-Jahn-Teller Effect and Magnetoelastic Coupling in Spin-Orbit Mott Insulators*, Phys. Rev. Lett. **122**, 057203 (2019).

[30] N. H. Sung, H. Gretarsson, D. Proepper, J. Porras, M. Le Tacon, A. V. Boris, B. Keimer, and B. J. Kim, *Crystal growth and intrinsic magnetic behaviour of* $Sr_2IrO_4$, Philosophical Magazine **96**, 413 (2016).

[31] J. P. Clancy, A. Lupascu, H. Gretarsson, Z. Islam, Y. F. Hu, D. Casa, C. S. Nelson, S. C. LaMarra, G. Cao, and Y.-J. Kim, *Dilute magnetism and spin-orbital percolation effects in* $Sr_2Ir_{1-x}Rh_xO_4$, Phys. Rev. B **89**, 054409 (2014).

[32] F. Ye, X. Wang, C. Hoffmann, J. Wang, S. Chi, M. Matsuda, B. C. Chakoumakos, J. A. Fernandez-Baca, and G. Cao, *Structure symmetry determination and magnetic evolution in* $Sr_2Ir_{1-x}Rh_xO_4$, Phys. Rev. B **92**, 201112 (2015).

[33] P. Liu, M. Reticcioli, B. Kim, A. Continenza, G. Kresse, D. D. Sarma, X.-Q. Chen, and C. Franchini, *Electron and hole doping in the relativistic Mott insulator* $Sr_2IrO_4$: *A first-principles study using band unfolding technique*, Phys. Rev. B **94**, 195145 (2016).

[34] N. Shannon, T. Momoi, and P. Sindzingre, *Nematic Order in Square Lattice Frustrated Ferromagnets*, Phys. Rev. Lett. **96**, 027213 (2006).

[35] Y. Iqbal, P. Ghosh, R. Narayanan, B. Kumar, J. Reuther, and R. Thomale, *Intertwined nematic orders in a frustrated ferromagnet*, Phys. Rev. B **94**, 224403 (2016).

[36] H. Zhang et al., *Comprehensive Electrical Control of Metamagnetic Transition of a Quasi-2D Antiferromagnet by In Situ Anisotropic Strain.* Advanced Materials **32**, 2002451 (2020).

[37] D. Kim, *Manipulation and Study of Emergent Phenomena in Oxide Ultrathin Films*, Ph.D. thesis, Seoul National University, 2024.

[38] K. Manna, G. Aslan-Cansever, A. Maljuk, S. Wurmehl, S. Seiro, and B. Büchner, *Flux growth of* $Sr_{n+1}Ir_nO_{3n+1}$ *(n = 1, 2, ∞) crystals*, Journal of Crystal Growth **540**, 125657 (2020).

[39] J. Kwon et al., *Spin-orbit coupling driven orbital-selective doping effect in* $Sr_2Ru_{1-x}Ir_xO_4$, Phys. Rev. B **103**, L081104 (2021).

[40] X. D. Zhu, R. Ullah, and V. Taufour, *Oblique-incidence Sagnac interferometric scanning microscope for studying magneto-optic effects of materials at low temperatures*, Review of Scientific Instruments **92**, 043706 (2021).

[41] A. Fried, M. Fejer, and A. Kapitulnik, *A scanning, all-fiber Sagnac interferometer for high resolution magneto-optic measurements at 820 nm*, Review of Scientific Instruments **85**, 103707 (2014).

[42] V. M. Katukuri, V. Yushankhai, L. Siurakshina, J. Van Den Brink, L. Hozoi, and I. Rousochatzakis, *Mechanism of Basal-Plane Antiferromagnetism in the Spin-Orbit Driven Iridate* $Ba_2IrO_4$, Phys. Rev. X **4**, 021051 (2014).

[43] T. Choi, Z. Zhang, H. Kim, S. Park, J. Kim, K. J. Lee, Z. Islam, U. Welp, S. H. Chang, and B. J. Kim, *Nanoscale Antiferromagnetic Domain Imaging using Full-Field Resonant X-ray Magnetic Diffraction Microscopy*, Advanced Materials **34**, 2200639 (2022).

# Supplementary Material for "Interplay of canted antiferromagnetism and nematic order in Mott insulating $Sr_2Ir_{1-x}Rh_xO_4$"

Hyeokjun Heo[1,2,*], Jeongha An[1,2,*], Junyoung Kwon[3], Kwangrae Kim[3], Youngoh Son[1], B. J. Kim[3], Joonho Jang[1,2,†]

[1]Department of Physics and Astronomy and Institute of Applied Physics, Seoul National University, Seoul 08826, South Korea

[2]Center for Correlated Electron Systems, Institute for Basic Science, Seoul 08826, South Korea

[3]Department of Physics, Pohang University of Science and Technology, Pohang 37673, South Korea

## S1. Magnetic responses at P2, P3, and P6 in undoped $Sr_2IrO_4$

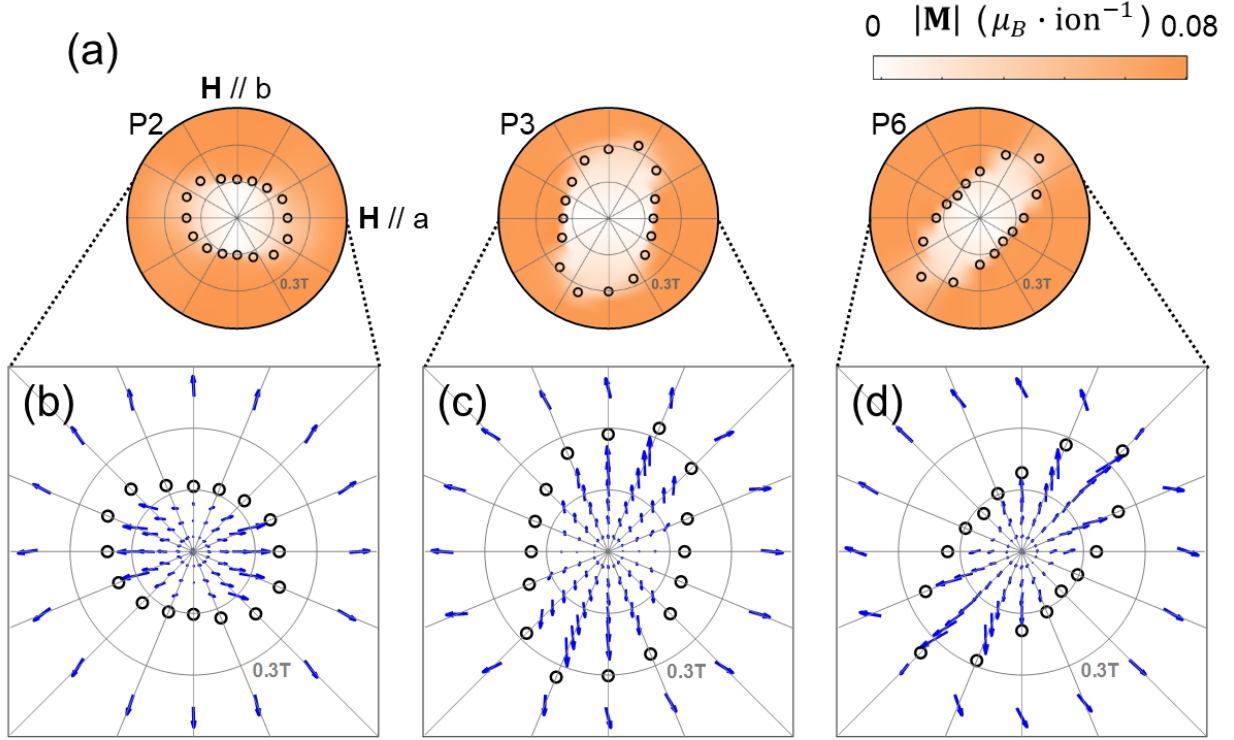

FIG. S1. Magnetic responses at P2, P3, and P6. (a) Magnetization magnitude |**M**| (color map) and $H_C$ (black circles) at P2, P3, and P6. (b)-(d) Detailed magnetic responses at (b) P2, (c) P3, and (d) P6. Critical fields $H_C$ are indicated by black circles, and the net magnetizations **M** are indicated by blue arrows. Note that the size of the arrows for $H < H_C$ is magnified four times for illustration.

---

* These authors contributed equally to this work.
† Contact author: joonho.jang@snu.ac.kr

## S2. Calculated magnetic responses for undoped $Sr_2IrO_4$

In **Fig. 3**, we present calculated magnetic responses at three representative positions; P1, P4, and P5. Comparing these results with the experimental data shown in the main text [**Figs. 2(d)-2(f)**], we confirmed that the model closely reproduces most of the key features.

We note that the ground state spin configuration at positions P1 and P4 follows the a-axis, as explained in the main text, and it is manifested as a positive value of the $K_{ab}$ term in the model calculation. When the directionality of the $H_C$ minima deviates from the a- or b-axis (positions P4 and P5), we introduce a finite value of the $N_{xy}$ term. Thus, following parameters are used in the model calculation. For P1, we used parameters of $J_{1c}$ = 13.0 μeV, $J_{2c}$ = -2.23 μeV, $\Delta_c$ = 0.02$J_{1c}$, $K_{ab}$ = 1.08 μeV, $\Gamma_1$ = 2.0 μeV, and zero otherwise. For P4, we used parameters of $J_{1c}$ = 8.62 μeV, $J_{2c}$ = -3.97 μeV, $\Delta_c$ = 0.02$J_{1c}$, $K_{ab}$ = 0.4 μeV, $\Gamma_1$ = 1.5 μeV and $N_{xy}$ = 1.24 μeV. For P5, we used parameters of $J_{1c}$ = 10.14 μeV, $J_{2c}$ = -3.12 μeV, $\Delta_c$ = 0.06$J_{1c}$, $\Gamma_1$ = 1.63 μeV and $N_{xy}$ = 0.81 μeV.

## S3. The $H_{xy}$ term with a phenomenological account of nematicity in the system

Here, we provide a *phenomenological* description of the magnetic response of a canted-AFM order coupled to an electronic subsystem with uniaxially-anisotropic susceptibility. The subsystem represents the electronic degrees of freedom that are not directly involved to form the pseudospins. We conjecture that the subsystem develops the anisotropy influenced by one of the nematic sources, such as a loop-current order and a spin-nematic quadrupolar order, that otherwise do not directly couple to the AFM spins.

The magnetic orders developed by the canted AFM moment of Ir spins, in general, can influence the spin polarization of the background electronic subsystem through an exchange interaction, and thus a free energy term that describes the induced polarization needs to be taken into account in the model Hamiltonian. Note that if the subsystem has an isotropic response, we expect the induced polarization energy would be either isotropic or follow the symmetry that is already present in the AFM phase, leading to a conventionally expected magnetic response of the AFM spins. However, additional nematic orders co-existing in the system—such as loop-current and d-wave spin-nematicity [1]—can lower the symmetry of the background system, leading to an anisotropic response pointing to unique directions.

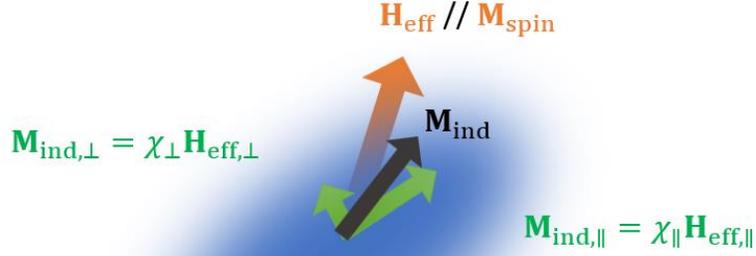

FIG. S2. Schematic picture depicting the influence of the net moment of the Ir spins (orange arrow) on the background electronic system (blue cloud) with nematic susceptibility. The angle developed by the $M_{spin}$ (or $H_{eff}$) and $M_{ind}$ provides the angular dependence of the energy of the system.

The spin exchange interaction energy between the Ir spins $S_{Ir}$ and the spins of the background subsystem can be approximated with an effective magnetic field generated by $S_{Ir}$ and an induced moment in the background subsystem as

$$E_{exch} = J \Sigma_n (\vec{S}_{Ir,n} \cdot \Sigma_m \vec{\sigma}_m) = -\vec{M}_{ind} \cdot \vec{H}_{eff}$$

, where $\vec{H}_{eff} = J\Sigma_n \vec{S}_{Ir,n} \approx JS\Sigma_i \cos\alpha_i$ is the effective field and $\vec{M}_{ind} = -\Sigma_m \vec{\sigma}_m$ is the induced moment (**Fig. S2**). Here, $n$ is the index for the summation of all Ir spins over a unit cell, $i$ is the index for each layer in a unit cell, and $m$ is the index for all electron spins in the subsystem. In the presence of an anisotropic response in the background subsystem, the induced moment can be further approximated to be $M_{ind} \approx \chi_{tot} \cdot \vec{H}_{eff} = \chi_{//} H_{eff,//} + \chi_\perp H_{eff,\perp}$, where $\chi_{tot}$ is the susceptibility tensor $(\chi_{//}, 0; 0, \chi_\perp)$ and $\chi_{//}$ and $\chi_\perp$ are parallel and perpendicular susceptibility factors, respectively. Then, the interaction energy between $S_{Ir}$ spins and the background spins reduces to

$$E_{exch} = -\vec{M}_{ind} \cdot \vec{H}_{eff} = -\chi_{tot} \cdot \vec{H}_{eff} \cdot \vec{H}_{eff}$$
$$= -\chi_{//} |\vec{H}_{eff} \cdot \hat{n}_{//}|^2 - \chi_\perp |\vec{H}_{eff} \cdot \hat{n}_\perp|^2$$
$$\approx -\chi_{//} (\Sigma_i \cos\alpha_i)^2 - \chi_\perp (\Sigma_i \sin\alpha_i)^2$$
$$= -\chi_{iso}\{(\Sigma_i \cos\alpha_i)^2 + (\Sigma_i \sin\alpha_i)^2\} - \chi_{ani}\{(\Sigma_i \cos\alpha_i)^2 - (\Sigma_i \sin\alpha_i)^2\}$$

, where $\chi_{iso} = (\chi_{//} + \chi_\perp)/2$ and $\chi_{ani} = (\chi_{//} - \chi_\perp)/2$ are isotropic and anisotropic susceptibility factors, respectively. Note that the isotropic term with $\chi_{iso}$ respects the symmetry already present in the system while the anisotropic term with $\chi_{ani}$ breaks additional symmetry not previously present in the system to generate unique two-fold rotational symmetry dictated by the supposed nematicity. Thus, we conclude that the anisotropic term is the minimal form that selectively distinguishes the nematic order.

After accounting for the 4-layer unit cell of $Sr_2Ir_{1-x}Rh_xO_4$, we used the following form in the model Hamiltonian to quantitatively describe the bond-directional electron nematicity:

$$-N_{xy}\left(\left(\frac{1}{4}\sum_{i=1}^{4} cos\alpha_i\right)^2 - \left(\frac{1}{4}\sum_{i=1}^{4} sin\alpha_i\right)^2\right),$$

where $\alpha_i$ is the angle of the layer moment in layer $i$.

### S4. Magnetic responses of the position P0 in undoped $Sr_2IrO_4$

We performed rotating-field MOKE measurements at position P0 (WFM region) of the undoped sample. **Figure S3** shows the experimental and calculated magnetization directions as a function of the magnetic field. For the description of the experimental results, we used parameters of $J_{1c}$ = -13.0 μeV, $J_{2c}$ = 3.0 μeV, $N_{xy}$ = 4.0 μeV and $E_{Th}$ = 0.6 μeV. It remains unclear whether the $\Gamma_2$ term is necessary in this case, given the limited magnetic field range of approximately 0.7 T.

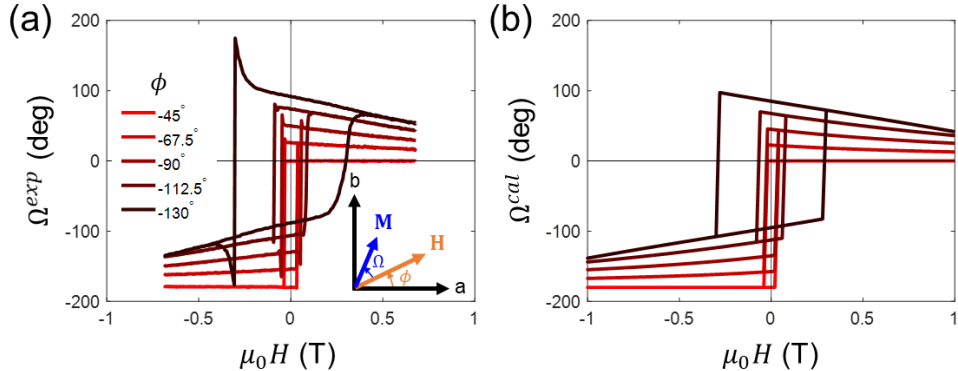

FIG. S3. MOKE measurement and model calculation of the WFM region (P0) in the undoped sample. (a) Measured magnetization directions $\Omega^{exp}$ for varying magnetic field directions $\phi$. The inset illustrates the geometry of the MOKE measurement, and the legend indicates $\phi$. (b) Calculated magnetization directions $\Omega^{cal}$ for varying magnetic field directions $\phi$. Note that $\Omega = tan^{-1}(M_T/M_L)$.

## S5. MOKE vs Temperature for various doping levels

In **Fig. S4**, we show the normalized longitudinal magnetization ($M_L$) as a function of temperature. All the data were collected while warming up. Except for position P6 of the undoped sample, Kerr angles were measured under ambient field conditions ($\mu_0 H < 0.005$ T). For position P6 of the undoped sample, we applied an in-plane magnetic field of $\mu_0 H = 0.2$ T because it exhibits a negligible Kerr angle in the zero field. We confirmed $T_N$ as 230 K (x = 0.00, position P6), 200 K (x = 0.00, position P0: WFM region), 170 K (x = 0.03), and 150 K (x = 0.05), which agrees with previously known values [2,3]. It is worth noting that, although the nematic order is thought to be present above $T_N$ [4-7], our Kerr measurements—which detect the magnetization from the canted AFM spin order—lose sensitivity above $T_N$ under the magnetic field range of our measurement. As a result, we were only able to confirm the presence of nematic order up to $T_N$.

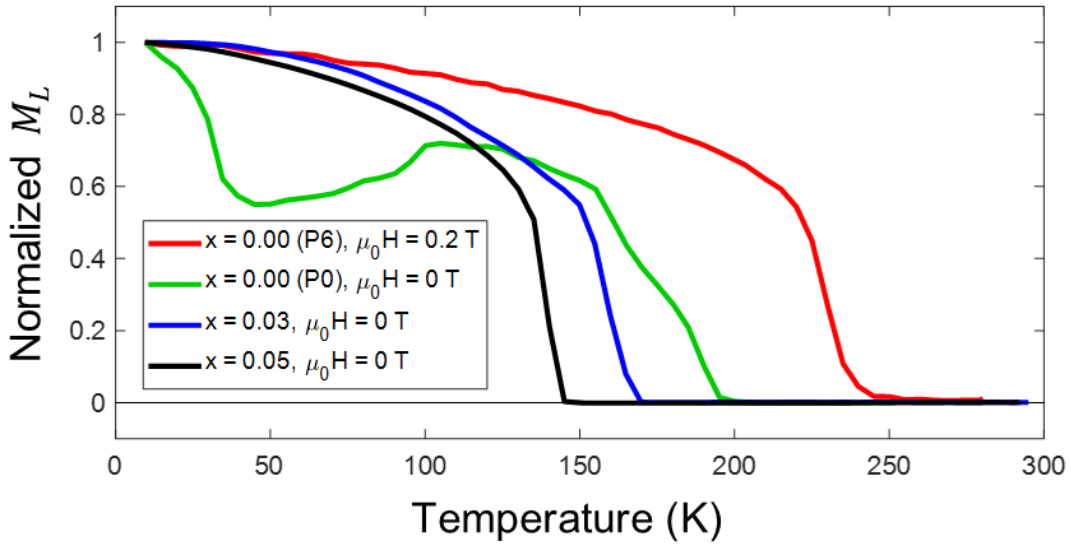

FIG. S4. Normalized longitudinal magnetization ($M_L$) as a function of temperature of $Sr_2Ir_{1-x}Rh_xO_4$. The doping levels and magnetic field values are indicated in the legend.

## S6. Differential susceptibility of undoped $Sr_2IrO_4$

In **Figs. 4(c)** and **4(d)** in the main text, we present images of two types of differential magnetic susceptibility: $\chi_{aa}^{H<H_c}$ and $\chi_{ba}^{H>H_c}$, with the magnetic field applied along the a-axis. In **Fig. 2(b)**, for example, $\chi_{aa}^{H<H_c}$ represents the slope of $M_L$ (black line) in the shaded region, while $\chi_{ba}^{H>H_c}$ represents that of $M_T$ (red line) outside the shaded region.

At zero magnetic field, two possible ground state configurations, $\bar{b}\bar{b}bb$ and $\bar{a}aa\bar{a}$, can exist, as shown in **Fig. 1(b)**. When the a-b symmetry breaking effect favors the a-axis, the $\bar{a}aa\bar{a}$ configuration becomes the only ground state. In this case, the magnetization remains zero for a small magnetic field applied along the a-axis, and thus $\chi_{aa}^{H<H_C}$ becomes zero. The magnetic behaviors at positions P1 and P4 are representative examples, as shown in **Figs. 2(d)** and **2(e)**. Otherwise, $\chi_{aa}^{H<H_C}$ shows a finite value, since applying a magnetic field along the a-axis tilts the moments in the $\bar{b}\bar{b}bb$ configuration.

For $H > H_C$, as all layer moments are flipped, the magnetization simply rotates to align with the external magnetic field, maintaining a constant magnitude. In regions with weak or no bond-directional order, the magnetization remains aligned along the a-axis, satisfying the given crystalline symmetry. However, when the bond-directional nematic order is present, the magnetic easy axis deviates from the crystal axis. If the order is along the x-axis (y-axis), the sample exhibits negative (positive) $M_T$. As the magnetic field increases, the absolute value of $M_T$ decreases, resulting in positive (negative) $\chi_{ba}^{H>H_C}$.

In **Fig. S5**, we present binary MOKE images to highlight the change in the domain distribution before and after the spin-flip transition, based on the data shown in **Figs. 4(c)** and **4(d)**. Given the distinct distributions of the orders, we argue that the two symmetry breaking effects arise from different physical origins.

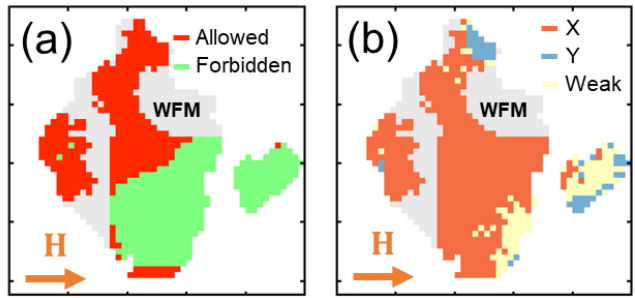

FIG. S5. Binary MOKE image of undoped $Sr_2IrO_4$. (a) Binary image of $\chi_{aa}$ for $H < H_C$. Red (green) color indicates where the $\bar{b}\bar{b}bb$ configuration is allowed (forbidden). (b) Binary image of $\chi_{ba}$ for $H > H_C$. Orange (sky) color corresponds to the X (Y) nematic domain. The yellow area indicates the region where $\chi_{ba}$ is insignificant to determine the type of the bond-directional order. The magnetic field is applied along the a-axis. Grey areas indicate WFM regions in the absence of an external magnetic field.

**S7. MOKE image and magnetic responses of the 5% Rh-doped sample**

We observed very similar magnetic distributions for both the x = 0.03 and x = 0.05 samples ($Sr_2Ir_{1-x}Rh_xO_4$). **Figure S6(a)** shows the MOKE image of the 5% Rh-doped sample. Similar to the observation in the 3% Rh-doped sample, we observed two different magnetic domains aligned along the x- or y-axis.
6

For the x = 0.05 sample, we measured the magnetic responses of both domains, denoted as X and Y in **Fig. S6(a)**. **Figures S6(b)-S6(e)** shows the measured and calculated magnetic responses for domain X. **Figures S6(f)-S6(i)** shows the measured and calculated magnetic responses for domain Y. For the model calculation of domain X, we used parameters of $J_{1c}$ = -13.0 μeV, $J_{2c}$ = 3.0 μeV, $\Gamma_2$ = 0.3 μeV, $N_{xy}$ = 0.4 μeV and $E_{Th}$ = 0.2 μeV. For the model calculation of domain Y, we used parameters of $J_{1c}$ = -13.0 μeV, $J_{2c}$ = 3.0 μeV, $N_{xy}$ = -0.8 μeV and $E_{Th}$ = 0.01 μeV. As expectedly, the negative sign of the $N_{xy}$ term is used to describe the magnetic behavior for domain Y.

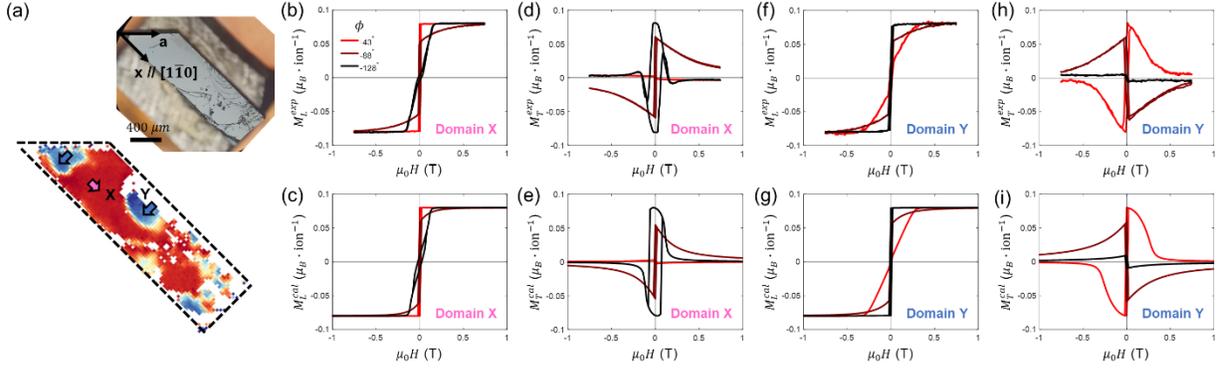

FIG. S6. MOKE measurement and model calculation of the 5% Rh-doped sample. (a) Scanning MOKE image at zero magnetic field, showing magnetizations aligned along the x-axis (denoted as X) or y-axis (denoted as Y). The inset presents an optical image of the sample. (b)-(e) Magnetic response of domain X: (b) measured $M_L$, (c) calculated $M_L$, (d) measured $M_T$ and (e) calculated $M_T$. (f)-(i) Magnetic response of domain Y: (f) measured $M_L$, (g) calculated $M_L$, (h) measured $M_T$ and (i) calculated $M_T$.

## S8. MOKE measurement and model calculation of the 3% Rh-doped sample

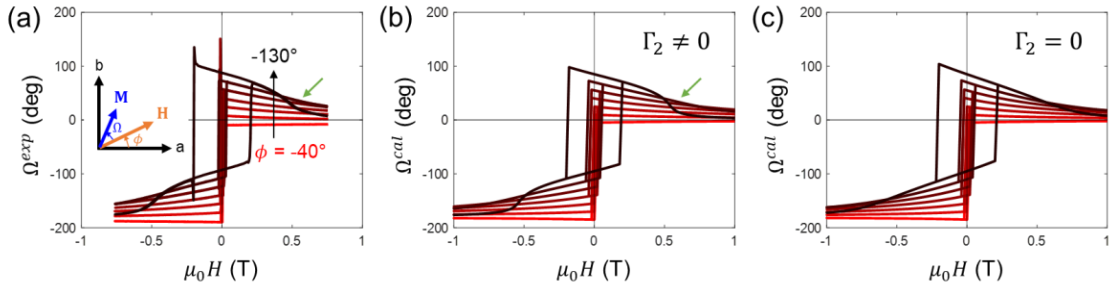

FIG. S7. MOKE measurement and model calculation of the 3% Rh-doped sample. (a) Measured magnetization directions $\Omega^{exp}$ for varying magnetic field directions $\phi$, ranging from -40° to -130° (in the order -40, -55, -70, -85, -100, -115 and -130°). The inset shows the geometry of the MOKE measurement. (b) Model calculation of magnetization directions $\Omega^{cal}$ with non-zero $\Gamma_2$ (=1.2 μeV). The kink in the $\phi$ = -130° data indicated by a green arrow is closely reproduced. (c) Model calculation with zero $\Gamma_2$, keeping all other parameters the same as in (b). Note that $\Omega = tan^{-1}(M_T/M_L)$.

## S9. Background subtraction procedure for the MOKE data

We subtract background signal from optical windows that is linearly proportional to the external magnetic field. **Figure S8(a)** shows nearly the same slope (μrad/T) for the $T = 5$ K and $T = 300$ K data, indicating that the background signal is nearly temperature-independent. The background-subtracted signal is plotted in **Fig. S8(b)**.

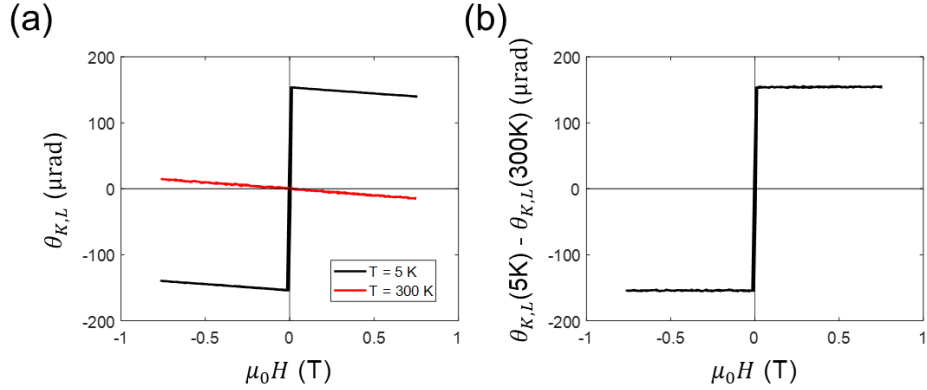

FIG. S8. Example of background subtraction. (a) Longitudinal Kerr angle at $T = 5$ K (black) and $T = 300$ K (red). (b) Background-subtracted longitudinal Kerr angle.

**Figure S9** shows the effect of background signal subtraction. As shown in **Fig. S9(a)**, the background signals have different slopes for different magnetic field directions. It comes from an experimental limitation where we rotated the sample instead of rotating the magnetic field. Due to a small misalignment of the c-axis of the sample with respect to the rotation axis, the incidence angle $\theta_{inc}$ varies for different sample rotations, resulting in different background signals. However, as shown in **Figs. S9(b)** and **S9(d)**, we obtain very similar results because the background signals are at least ~10 times smaller than the sample signal itself, confirming that this misalignment is negligible for the interpretation of our data.

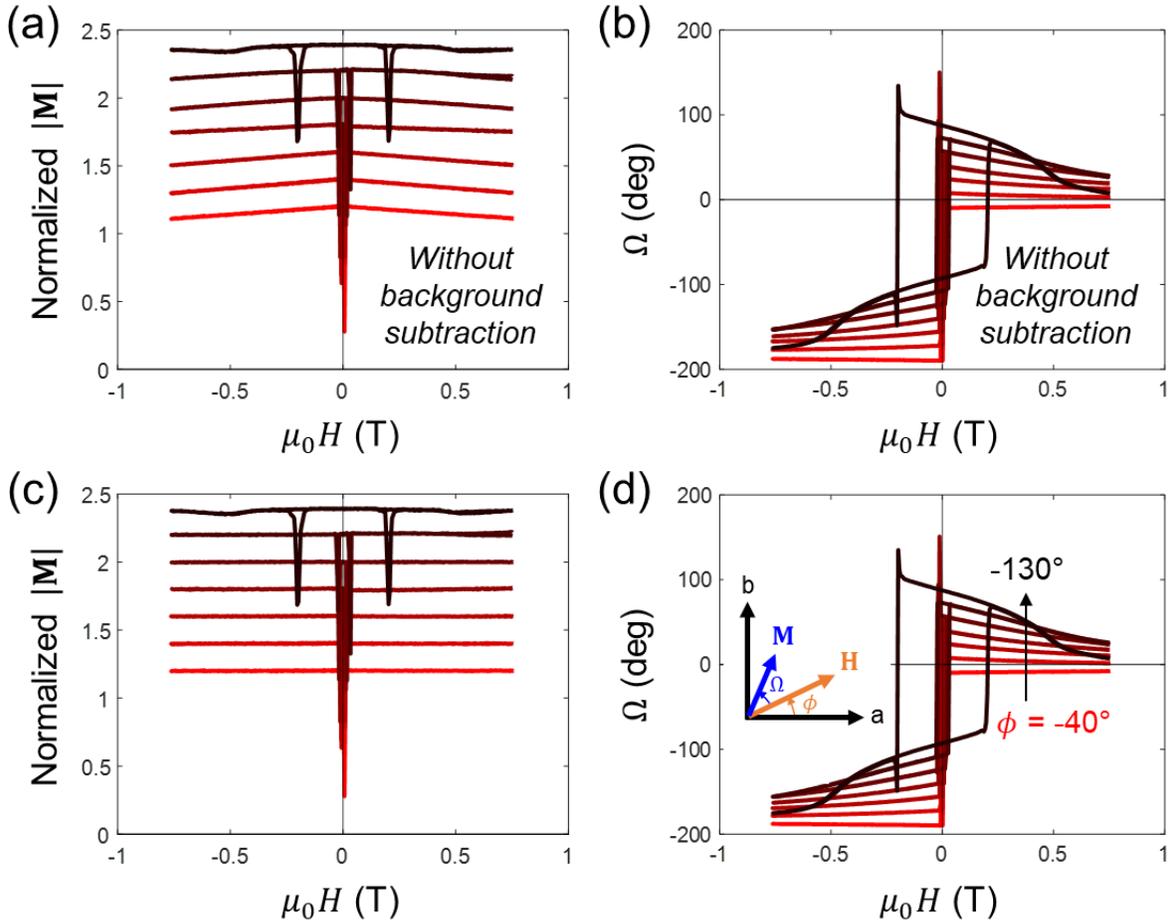

FIG. S9. Comparison between background-subtracted data and raw data. (a),(b) Raw data of (a) magnetization magnitudes |**M**| and (b) magnetization directions Ω. (c),(d) Background-subtracted (c) magnetization magnitudes |**M**| and (d) magnetization directions Ω. The Y-values in (a) and (c) are vertically shifted for clarity.

## S10. Table for the various magnetic ground states and its energy

In **Table SI**, various magnetic configurations and their energies are listed in zero magnetic field. The energies are calculated with the Hamiltonian $H = H_{inter} + H_{JT}$ from Eqs. 4, 5. For example, the $\bar{b}\bar{b}bb$ and $\bar{a}aa\bar{a}$ magnetic patterns have zero net magnetization with an energy of $-j_{2c} -\delta_c -\gamma_1$ and become degenerate ground states. However, in the presence of uniaxial anisotropy $K_{ab}$ with positive (negative) sign, $\bar{a}aa\bar{a}$ ($\bar{b}\bar{b}bb$) becomes the only ground state.

| Energy term | Pattern ($M = 0$) | | | | | | | Pattern ($M \neq 0$) | | | |
|---|---|---|---|---|---|---|---|---|---|---|---|
| | $\bar{b}\bar{b}bb$ $\bar{a}aa\bar{a}$ | $b\bar{b}\bar{b}b$ $\bar{a}\bar{a}aa$ | $\bar{b}bb\bar{b}$ $\bar{a}a\bar{a}a$ | $\bar{a}a\bar{b}b$ $\bar{b}\bar{a}ab$ $\bar{b}b\bar{a}a$ $\bar{a}\bar{b}ba$ | $a\bar{a}\bar{b}b$ $\bar{b}a\bar{a}b$ $bb\bar{a}a$ $\bar{a}\bar{b}ba$ | $\bar{a}\bar{b}ab$ $\bar{b}\bar{a}ba$ | $a\bar{b}\bar{a}b$ $b\bar{a}\bar{b}a$ | $bbbb$ $aaaa$ | $aabb$ $b\bar{a}a\bar{b}$ $bbaa$ $a\bar{b}\bar{b}a$ | $baab$ $\bar{b}\bar{b}aa$ $abba$ $\bar{a}\bar{a}bb$ | $abab$ $\bar{a}b\bar{a}b$ $baba$ $\bar{b}a\bar{b}a$ |
| $J_{1c}$ | | | -1 | -1/2 | -1/2 | | | +1 | +1/2 | +1/2 | |
| $J_{2c}$ | -1 | -1 | +1 | | | -1 | -1 | +1 | | | +1 |
| $\delta_c$ | -1 | +1 | | | | | | | | | |
| $\gamma_1$ | -1 | -1 | -1 | | | | | -1 | | | |
| $\gamma_2$ | | | | | | | | | | | |
| Energy term | Pattern ($M = 0$) | | | | | | | Pattern ($M \neq 0$) | | | |
| | $xx\bar{x}\bar{x}$ $y\bar{y}\bar{y}y$ | $x\bar{x}\bar{x}x$ $yy\bar{y}\bar{y}$ | $x\bar{x}x\bar{x}$ $y\bar{y}y\bar{y}$ | $x\bar{x}\bar{y}y$ $x\bar{y}y\bar{x}$ $y\bar{y}\bar{x}x$ $y\bar{x}x\bar{y}$ | $x\bar{x}y\bar{y}$ $x\bar{y}y\bar{x}$ $y\bar{y}x\bar{x}$ $yx\bar{x}\bar{y}$ | $xy\bar{x}\bar{y}$ $yx\bar{x}\bar{y}$ | $x\bar{y}\bar{x}y$ $y\bar{x}\bar{y}x$ | $xxxx$ $yyyy$ | $xxyy$ $x\bar{y}\bar{y}x$ $yyxx$ $\bar{y}xx\bar{y}$ | $xyyx$ $\bar{y}\bar{y}xx$ $yxxy$ $xx\bar{y}\bar{y}$ | $xyxy$ $x\bar{y}x\bar{y}$ $y\bar{x}y\bar{x}$ $yxyx$ |
| $J_{1c}$ | | | -1 | -1/2 | -1/2 | | | +1 | +1/2 | +1/2 | |
| $J_{2c}$ | -1 | -1 | +1 | | | -1 | -1 | +1 | | | +1 |
| $\delta_c$ | | | | +1/2 | -1/2 | -1 | +1 | | +1/2 | -1/2 | |
| $\gamma_1$ | | | | | | | | | | | |
| $\gamma_2$ | -1 | -1 | -1 | | | | | -1 | | | |

TABLE SI. Magnetic configurations and corresponding energies in zero magnetic field, using the Hamiltonian $H = H_{\text{inter}} + H_{JT}$ from Eqs. 4, 5.

## SUPPLEMENTARY REFERENCES

[1] Y. Iqbal, P. Ghosh, R. Narayanan, B. Kumar, J. Reuther, and R. Thomale, *Intertwined nematic orders in a frustrated ferromagnet*, Phys. Rev. B **94**, 224403 (2016).

[2] N. H. Sung, H. Gretarsson, D. Proepper, J. Porras, M. Le Tacon, A. V. Boris, B. Keimer, and B. J. Kim, *Crystal growth and intrinsic magnetic behaviour of* $Sr_2IrO_4$, Philosophical Magazine **96**, 413 (2016).

[3] C. Lu and J. Liu, *The* $J_{eff} = 1/2$ *Antiferromagnet* $Sr_2IrO_4$: *A Golden Avenue toward New Physics and Functions*, Advanced Materials **32**, 1904508 (2020).

[4] L. Zhao, D. H. Torchinsky, H. Chu, V. Ivanov, R. Lifshitz, R. Flint, T. Qi, G. Cao, and D. Hsieh, *Evidence of an odd-parity hidden order in a spin–orbit coupled correlated iridate*, Nat. Phys. **12**, 32 (2016).

[5] J. Jeong, Y. Sidis, A. Louat, V. Brouet, and P. Bourges, *Time-reversal symmetry breaking hidden order in* $Sr_2(Ir,Rh)O_4$, Nat. Commun. **8**, 15119 (2017).

[6] H. Murayama et al., *Bond Directional Anapole Order in a Spin-Orbit Coupled Mott Insulator* $Sr_2(Ir_{1-x}Rh_x)O_4$, Phys. Rev. X **11**, 011021 (2021).

[7] H. Kim et al., *Quantum spin nematic phase in a square-lattice iridate*, Nature **625**, 264 (2024).